\documentclass[11pt]{article}
\newcommand{\ct}[1]{$^{\cite{#1}}$}

\newcommand{\bee}{\begin{equation}}
\newcommand{\ee}{\end{equation}}
\newcommand{\beea}{\begin{eqnarray}}
\newcommand{\eea}{\end{eqnarray}}

\begin{document}
\title{Independent Operators at Different Dimension}
\author{{Yin Chen$^1$, Daqing Liu$^1$, Yubin Liu$^2$, Jimin Wu$^1$}\\
        {\small $^1$Institute of High Energy Physics, Chinese Academy
                of Sciences, P.R. China}\\
        {\small $^2$Nankai university, P.R. China}\\
       }
\maketitle
\begin{minipage}{5.5in}

\vskip 0.2in {\bf Abstract}\\
To apply lattice QCD in the calculation of glueball spectrum it is
needed firstly to know associated operators acting on vacuum. We
show how to find all the independent representations and
operators, of group $SO(3)^{PC}$ at different dimension, since the
work is not trivial. Then, we decompose these representation into
irreducible representation of $O^{PC}$ group, which are listed in
the note. At last we argue that $f_J(2220)$ and $g_T$ states can
not be tensor glueball simultaneously.
\end{minipage}

\vskip 0.3in \indent

\newpage

This is a note when we calculate the glueball spectrum in quenched
lattice QCD. We are focusing on the independent currents at each
dimension here.

As we know,  in path quantization, an arbitrary bounded particle,
especially, glueball, with $J^{PC}$, where $J$ is spin of
glueball, or total angular momentum of the composite particle, $P$
and $C$ are quantum number of parity and charge conjugate
respectively, can be considered as a wave functional, which is an
eigen-functional of Hamiltonian in QCD. With respect to the basis
of eigenvalue of canonical fields, $A_i=A_i^aT_a$ and therefore
$B_i=B_i^aT^a={1\over 2}\epsilon_{ijk}F_{jk}={1\over
2}\epsilon_{ijk}(\partial_jA_k-\partial_kA_j-i [A_j,A_k])$, a
state wave functional $|\psi>$, particularly, glueball state with
$J^{PC}$, is generated\ct{ldq1} by corresponding current acting on
the vacuum functional $|0>$, \bee |\psi>=O|0>, \ee where $O$ is
gauge invariant operator with $J^{PC}$. In free QED, one can
calculate operator $O$ for each state. However, in QCD, one can
not, due to the non-perturbational and nonlinear behavior of QCD.
Operator $O$ can be written as

\bee O=\sum\limits_{n\geq 4}\sum\limits_{c^n_i}c^n_i O^{(n)}_i,
\ee where $O^{(n)}_i$ is the \(i\)'th operator with $J^{PC}$ in
dimension $n$, and $c^n_i$ is coefficient to solve. Generally, if
we set single or only finite $c^n_i$ nonvanishing, operator $O$
acting on vacuum functional will generate infinite eigenstates
(physical states with definite energy) with the same $J^{PC}$, or
vice versa, there are infinite nonvanishing $c^n_i$ for each
eigenstate.

Therefore, to extract a definite glueball, one needs infinite
operators $O^{(n)}_i$'s and calculates infinite coefficients
$c^n_i$'s. This is an impossible mission, since in practice we can
only use finite operators $O^{(n)}_i$'s. However, thanks to
functional analysis\ct{prd99,prd97}, using these finite operators
we can always obtain optimal approximation to the ultimate
operator $O$ for definite state. Although eigen-currents under
optimal approximation also generate infinite states, we expect the
projections of these currents on heavier states are small and we
can neglect contributions of those heavier states.

According to functional analysis, to extract a definite glueball,
for example, quintet states $2^{++}$, we take such steps. Step
one, one writes independent $2^{++}$ operators dimension by
dimension. In practice, the operators are surely truncated at some
dimension, for instance, dimension seven or dimension eight. Step
two, since we are working on lattice, the rotation symmetry is
broken into cubic symmetry. We thus decompose this operators into
different irreducible representation of $O^{PC}$ group. For
instance, we decompose $2^{++}$ operator into $E^{++}$ and
$T_2^{++}$ representation. Step three, we represent these
operators on lattice by Wilson loops, the continuum limits of
which are just the corresponding operators. Step four, for
instance, for those nonequivalent $E^{++}$ irreducible
representations belonging to different $2^{++}$ representations in
$SO(3)^{PC}$ group, we use variation principle to extract optimal
approximations to different glueball $2^{++}$, at finite lattice
spacing $a$. The last step, of course, we extrapolate the results
into continuum limit.

This method has constraints. If we use three independent
operators, which belong to three different representations of
$2^{++}$, we at most extract three different $2^{++}$ glueballs.
Secondly, these operators should be independent, otherwise we
shall meet singularity in the utilizing of variation principle.
Therefore, to find all the independent currents at each dimension
is important on lattice, since the complete basis of glueball
states is generated by all the gauge independent currents, which,
as shown, are made up of chromomagnetic fields, $B_i$'s, and
covariant derivatives, $D_i$'s, where subscripts stand for
directions and superscripts are color indices.  This is just the
topic of this note.

On one hand, for static glueball states one should sum operators
over all spatial points on lattice to construct corresponding
current. On the other hand, these operators should be local and
gauge invariant products of $B_i$ and $D_i$ at the same time. For
instance, $\sum\limits_{\bf x} B_i^aB^a_i$ generates $0^{++}$
(infinite) glueballs when it acts on vacuum (The summation over
spatial points will be suppressed thereinafter).

For this aim  we first need to know the behavior of $B_i^a$ and
$D_i^{bc}$ under rotation. The rotation transformation is divided
in two parts, one is the rotation in orbital space, and the other
is the rotation in spin space. $B_i^a$ is vector operator under
rotation transformation in spin space and scalar operator under
rotation transformation in orbital space. Therefore, operator
$B_i^a$ can be interpreted as spinor with unit spin. Whereas,
$D_i^{bc}=\partial_i\delta^{bc}-iA_i^at^a$, where
$(t^b)_{ac}=if^{abc}$ is generator in adjoint representation, is
the vector operator only under total rotation transformation. This
is because that the first term in covariant operator,
$\partial_i\delta^{bc}$, is vector under orbital rotation and
scalar under spin rotation, while the second term in covariant
operator, $-iA_i^at^a$, has a reverse character,  for it is scalar
under orbital rotation and vector under spin rotation. Thus, the
covariant derivative operator can not be interpreted as orbital
angular momentum vector which ascends or descends unit orbital
angular momentum. But, if we ignore the term $-iA_i^at^a$, it can.
This is nearly true if the glueball is sufficient small and gauge
fields are smooth enough, for one can always let $A_i^a(x)\equiv
0$ at one point, for instance the center of glueball mass, by
performing a local gauge transformation. Attributed to this reason
we sometimes call covariant derivative as orbital vector, although
this will take many ambiguities.

It seems firstly that finding independent current at each
dimension is a trivial mission. However, due to the so-called
Bianchi identity, one always meets some subtleties. We shall show
how to treat these subtleties by an example, {\it i.e.} finding
all the independent operators at the case of dimension six. Such
treatment can be generalized into other dimension directly.

All the operators at dimension six can be categorized into two
types: type one is as $Tr((D_iD_jB_k)B_l)$ and type two is as
$Tr(B_iB_jB_k)$. $Tr((D_iB_j)(D_kB_l))$ is not an independent
type, for operators in such type can be rewritten as the form of
type one up to a complete covariant derivative, which will be
discarded at last.

We first consider operators in type one. Taking advantage of
standard basis of chromomagnetic fields $B_\pm=\mp (B_1\pm
iB_2)/\sqrt{2}$ and $B_0=B_3$, we find the total spin of $B_kB_l$
can be 0,\,1,\,2. Similarly, the standard basis $D_\pm=\mp (D_1\pm
iD_2)/\sqrt{2}$ and $D_0=D_3$ makes that the total 'orbital
angular momentum' of operators $D_iD_j$ be 0,\,1,\,2. It seems
that there are 81 operators. But, this is not true.

First, we rewritten such operator using a technique of
 symmetry-antisymmetry decomposition, \beea
Tr((D_iD_jB_k)B_l)&=&{1\over
2}Tr\{[(D_{ij}^sB_k)B_l+(D_{ij}^sB_l)B_k]+[(D_{ij}^sB_k)B_l-(D_{ij}^sB_l)B_k]
\nonumber \\ &&+
[(D_{ij}^aB_k)B_l+(D_{ij}^aB_l)B_k]+[(D_{ij}^aB_k)B_l-(D_{ij}^aB_l)B_k]
\},\label{decompose}\eea where $D_{ij}^s={1\over
2}(D_iD_j+D_jD_i)$ and $D_{ij}^as={1\over 2}(D_iD_j-D_jD_i)$. This
technique is crucial. It reduces all the operators into the forms
of symmetry, antisymmetry and admixture symmetry-antisymmetry. In
each form all the operators at the same dimension make up of an
invariant space of group $SO(3)^{PC}$. Eq. (\ref{decompose}) tells
us that operators in type one are divided into four forms. But,
terms in the last two square bracket are in fact operators in type
two because of the identity $[D_i,D_j]=[F_{ij},\cdot]$. (There is
in fact only one independent nonvanishing operator
$\epsilon_{ijk}Tr(B_iB_jB_k)$ in the form
$Tr((D_{ij}^asB_k)B_l)$.) It is easy to check the operator in the
second square bracket is vanishing up to a complete covariant
derivative. Thus, for operator belonging to type one it is enough
to consider the operators as the form shown in the first quare
bracket. Therefore, for operator of type one we only consider the
case that the total spin and total orbital angular momentum are 0
or 2. There are total 36 operators to be considered now.

The $J^{PC}$ of the total operators in type one  are summarized in
table 1, then,

\begin{table}[h]
\begin{center}
\begin{tabular}{|c|c|c|}
\hline
  spin $\backslash$ orbital & 0 & 2 \\ \hline
  0 & $0^{++}(\, ^1S_0)$ & $2^{++}(\, ^1D_2)$  \\ \hline
  2 & $2^{++}(\, ^5S_2)$ & $0^{++}(\,^5D_0),\, 1^{++}(\,^5D_1),
  \,2^{++}(\,^5D_2),\,3^{++}(\,^5D_3),\,4^{++}(\,^5D_4)$ \\
  \hline
  \end{tabular}
  \end{center}
  \caption {The $J^{PC}$ of the total operators in type one. The state
  symbols in bracket are obtained by interpreting covariant
  derivative as orbital angular momentum vector. Surely such
  interpretation has ambiguities.
   }
\end{table}

Now we turn to operators in type two. type two can be divided into
two forms under symmetry-antisymmetry decomposition,
$Tr(B_i(B_jB_k-B_kB_j))$ and $Tr(B_i(B_jB_k+B_kB_j))$. In the
first form there is only one independent operator,
$\epsilon_{ijk}Tr(B_iB_jB_k)$, which is $0^{++}$. This operator
can also be regarded as the  one in type one, as shown before. We
pay attention on the second form here, the $PC$ of which is $+-$.
For the case $i\neq j\neq k$, there is one independent operator,
$Tr(B_1B_2B_3+B_1B_3B_2)$, which belongs to $A_2^{+-}$ in cubic
group and therefore belongs to $3^{+-}$ in $SO(3)^{PC}$ group
concerning to the total number of operators, ten, in form two.
There are three operators at the case $i=j=k$, which construct to
the representation $T_1^{+-}$. For the case $i\neq j=k$, there are
six operators. Therefore, there are total eleven independent
operators in type two. As argued, one of them belongs to $0^{++}$,
and seven of them belong to $3^{++}$. The remnant three operators
belong to $1^{+-}$, since three operators
$Tr(B_i(B_1B_1+B_2B_2+B_3B_3))$'s make up of the representation
$1^{+-}$ of the $SO(3)^{PC}$ group.

We then have operators belong to three nonequivalent
representations $0^{++}$ and $2^{++}$, one representation
$1^{++}$, $3^{++}$, $4^{++}$, $1^{+-}$ and $3^{+-}$ respectively.

It seems the topic is finished, but the game is not so simple, due
to the Bianchi identity, $I=D_1B_1+D_2B_2+D_3B_3\equiv 0$.

Now we consider the constraint introduced by Bianchi identity.
Bianchi identity is a scalar operator. Then, the operators
containing Bianchi identity at dimension six are as the form
$Tr(B_iD_jI)$, {\it i.e.} they belong to type one and therefore,
their $PC$ are $++$. Since $B_iD_j$ can compose representation
$J=0$, 1 and 2 under rotation transformation, we conclude that
operators containing Bianchi identity can compose representations
$0^{++}$, $1^{++}$ and $2^{++}$ under $SO(3)^{PC}$ group. We
emphasize here that for general case, we need also use
symmetry-antisymmetry decomposition to study $J^{PC}$'s of the
operators containing Bianchi identity, although at the case of
dimension six or five we need not.

Therefore, operators belonging to $1^{++}$ are vanishing. The
three representations $0^{++}$ are not independent and there are
only two independent nonequivalent representations $0^{++}$. It is
similar for $2^{++}$. For two independent representations
$0^{++}$, we pick on states $^1S_0$ and
$\epsilon_{ijk}Tr(B_iB_jB_k)$, while for $2^{++}$, states $^1D_2$
and $^5S_2$ respectively.

This statement can be checked by a directly calculation. For
instance, suppose $\alpha_2$ is the operator of state $^5D_0$ up
to a factor, \beea \alpha_2 &=& Tr\{
-B_1(-2D_1^2+D_2^2+D_3^2)B_1-B_2(-2D_2^2+D_1^2+D_3^2)B_2-
B_3(-2D_3^2+D_1^2+D_2^2)B_3 \nonumber \\ &&
+3B_2(D_2D_1+D_1D_2)B_1+3B_1(D_3D_1+D_1D_3)B_3+3B_2(D_3D_2+D_2D_3)B_3\},
\eea and $\alpha_0$ is the operator of state $^1S_0$ up to a
factor, \bee  \alpha_0=Tr\{B_1(D_1^2+D_2^2+D_3^2)B_1+
B_2(D_1^2+D_2^2+D_3^2)B_2+ B_3(D_1^2+D_2^2+D_3^2)B_3\}, \ee one
has $\alpha_2+\alpha_0\equiv 0$ if he takes advantage of Bianchi
identity.

The analysis for operators at dimension six can be generalized
into other dimension. For instance, operators at dimension 5
belong to representations $0^{-+}$ and $2^{-+}$ respectively, and
operators at dimension 4 belong to representations $0^{++}$ and
$2^{++}$ respectively.Operators at dimension five has no
nonvanishing representation $1^{-+}$ due to Bianchi identity.

In summary, all the operators at dimension four, five and six  can
be categorized into three independent representations $0^{++}$ and
$2^{++}$, one representation $3^{++}$, $4^{++}$, $1^{+-}$,
$3^{+-}$, $0^{-+}$ and one representation $2^{-+}$. One can make
use of C-G coefficients to obtain these operators.

These representations are irreducible in $SO(3)^{PC}$ group, but
they are reducible representations in group $O^{PC}$. Since on
lattice we only have $O^{PC}$ symmetry, we shall furthermore
reduce these representations into irreducible representations of
group $O^{PC}$. The results are very lengthy, and we only list
some of them, which will be used in our lattice simulations. In
each irreducible representation only one operator is shown. This
is enough even for representations $E^{PC}$, $T_1^{PC}$ and
$T_2^{PC}$, because as long as one operator is found, other
operator(s) belonging to the same representation can easily be
obtained by suitable cubic transformations.

\begin{itemize}
    \item Three $0^{++}$ representations:
    \begin{enumerate}
        \item The first representation $0^{++}$, which is of four dimension,
        \bee a_1 \propto Tr(B_1^2+B_2^2+B_3^2); \ee
        \item The second representation $0^{++}$, which is of six dimension,
        \bee  a_1 \propto Tr[B_1(D_1^2+D_2^2+D_3^2)B_1+B_2(D_1^{2}+D_2^{2}+D_3^{2})B_2
+B_3(D_1^{2}+D_2^{2}+D_3^{2})B_3]; \ee
        \item The third representation $0^{++}$, which is of six dimension,
        \bee a_1\propto Tr[B_1(B_2B_3-B_3B_2)]. \ee
    \end{enumerate}

    \item Three $2^{++}$ representations:
        \begin{enumerate}
        \item The first representation $2^{++}$, which is of four dimension,
        is decomposed into irreducible representation $E^{++}$ and
        $T_2^{++}$.  For representation $E^{++}$,
        \bee e^1 \propto Tr(B_1^2-B_2^2), \ee
        while for $T_2^{++}$,
        \bee t_2^1\propto Tr(B_2B_3); \ee
        \item The second representation $2^{++}$ is of six dimension.
        For representation $E^{++}$,
        \bee e^1\propto Tr[B_1(D_1^2+D_2^2+D_3^2)B_1
        -B_2(D_1^{2}+D_2^{2}+D_3^{2})B_2], \ee
        while for $T_2^{++}$,
        \bee t_2^1\propto Tr[B_2(D_1^2+D_2^2+D_3^2)B_3] ; \ee
        \item The third representation $2^{++}$ is of six
        dimension. For representation $E^{++}$,
        \bee e^1\propto Tr[B_1(D_1^{2}-D_2^{2})B_1
        +B_2(D_1^{2}-D_2^{2})B_2+B_3(D_1^{2}-D_2^{2})B_3], \ee
        while for $T_2^{++}$,
        \bee t_2^1\propto Tr[B_1D_3D_2B_1+B_2D_3D_2B_2+B_3D_3D_2B_3] . \ee
    \end{enumerate}
    \item Representation $4^{++}$ only occurs in dimension six.
    This representation is reduced into $A_1^{++}$, $E^{++}$ and
    two $T_2^{++}$'s. For representation $A_1^{++}$,
        \beea
        a_1&\propto& Tr[B_1(2D_1^{2}-D_2^{2}-D_3^{2})B_1
        +B_2(-D_1^{2}+2D_2^{2}-D_3^{2})B_2 \nonumber \\
        &&+B_3(-D_1^{2}-D_2^{2}+2D_3^{2})B_3-2B_2(D_2D_1+D_1D_2)B_1
        \nonumber \\
        &&-2B_1(D_3D_1+D_1D_3)B_3-2B_2(D_3D_2+D_2D_3)B_3],
        \eea
    and for representation $E^{++}$
        \beea
        e^1&\propto&Tr[B_1(D_1^{2}-D_3^{2})B_1
        -B_2(D_2^{2}-D_3^{2})B_2-B_3(D_1^{2}-D_2^{2})B_3  \nonumber \\
        &&-4B_1D_3D_1B_3+4B_2D_3D_2B_3].
        \eea
    \item Representation $3^{++}$,  which is reduced into $A_2^{++}$, $T_1^{++}$ and
    $T_2^{++}$, only occurs in dimension six. For representation $A_2^{++}$,
        \bee
        a_2\propto Tr[B_1(D_2^{2}-D_3^{2})B_1
        +B_2(D_3^{2}-D_1^{2})B_2 +B_3(D_1^{2}-D_2^{2})B_3],
        \ee
    and for representation $T_1^{++}$,
        \beea
        t_1^1&\propto&Tr[-B_2D_2D_3B_2+B_3D_2D_3B_3
        +2B_2D_1D_3B_1  \nonumber \\
        &&-2B_1(D_2D_1+D_1D_2)B_3+B_2(D_2^2-D_3^2)B_3].
        \eea
    \item We meet representation $0^{-+}$ in dimension 5,
        \bee a_1\propto Tr[B_2D_1B_3+B_3D_2B_1+B_1D_3B_2]. \ee
    \item Representation $2^{-+}$ is met in dimension 5. For
    $E^{ -+}$,
        \bee e^1\propto Tr[B_2D_1B_3+B_1D_2B_3], \ee
    and for $T_2^{-+}$
        \bee t_2^1\propto Tr[B_1D_3B_3+B_2D_2B_1]. \ee
    \item Representation $1^{+-}$ at dimension 6 is reduced into
    irreducible representation $T_1^{+-}$,
        \bee t_1^1\propto Tr[B_1B_1^2+B_1B_2^2+B_1B_3^2]. \ee
    \item Representation $3^{+-}$ occurs in dimension six. We
    show the operator in $A_2^{+-}$,
        \bee a_2\propto Tr[B_2B_1B_3+B_2B_3B_1]. \ee
\end{itemize}

Now we have listed the results of decomposing twelve independent
representations into irreducible representations in $O^{PC}$
group. If we use variation principle we obtain twelve glueballs.
these glueballs, except excited scalar and tensor glueballs, have
been measured by other authors. The listed results tell us the
structures of different glueballs, which is useful, for instance,
in glueball decay or in condensate of different current in
glueballs.

Although we do not make any lattice simulation here, we can obtain
some physical conclusions from the results, one of which will be
shown in the next paragraph.

to apply variation principle we can obtain 12 glueballs, which are
in fact the optimal approximation of the real glueballs.
Attributed to dimension analysis, it is reasonable to regard
$4^{++}$ glueball is the heaviest glueball with $PC=++$ among
those 12 glueballs. As pointed out by previous works\ct{ldq2}, the
mass of which is about $m_{4^{++}}=3\sim 4Gev$. We have shown
elsewhere that it is enough to use chromomagnetic fields and their
covariant derivative in the calculation of glueball spectrum
merely. Therefore, there are three and only three $2^{++}$ states
lower than $m_{4^{++}}$. One possibly debates that their masses
will be corrected by correction from fermion-loop and from
higher-dimension operators. On one hand, to consider the
correction of fermion-loops we should go into full QCD. However,
it is believed that the quenched approximation almost does not
change the ratio between different states. We think the mass
correction from fermion-loop is mainly on scale shift. On the
other hand, we shall get decreased glueball masses on lattice if
we consider the correction from operators at higher dimension.
Such decrease is both on $4^{++}$ glueball and on $2^{++}$
glueballs. We conclude that mainly effect of higher dimension
operator is also a scale shift. Therefore, the statement, there
are three and only three $2^{++}$ states lower than $m_{4^{++}}$,
is right or almost right even including the two corrections,
possibly up to a scale shift. If $f_J(2220)$ (or $\xi(2230)$) is a
tensor glueball, one must find other two tensor glueballs with
mass lower than $m_{4^{++}}$. But, since there are only three
tensor glueballs below $m_{4^{++}}$, it is difficult to place
three $g_T$ states, which are more like tensor glueballs. However,
if $g_T$ are interpreted as three glueballs, $f_J(2220)$ should
never be tensor glueball. The tensor glueball interpretations of
$f_J(2220)$ and the tensor glueball interpretation of $g_T$ are
conflictive interpretations, therefore. Since there are more
supports of the tensor glueball interpretation of $g_T$, at least
in nowadays, $f_J(2220)$ can not be tensor glueball, if it exists.

There are quenched lattice gauge calculation supporting the tensor
glueball interpretation of $f_J(2220)$, for instance, reference
\cite{prd99} claims that masses of ground scalar and ground tensor
glueball are $1.73Gev$ and $2.40Gev$ respectively. The scale
parameter is determined by phenomenal Sommer potential\ct{sommer}
in these works\ct{prd99,prd97}, that is , $r_0$ is determined by
$r^2\frac{dV}{dr}|_{r=0.49fm}=1.65$, since there is no exclusive
experiment result on glueball mass. However, $r_0$ dependents on
different phenomenal potential models, for instance, as author
pointed out in reference \cite{sommer}, the logarithmic potential
gives values that are lower by about 10\%. Combining the argument
in the last paragraph, we think that there may exist a global
scale shift in these works\ct{prd99,prd97}. If this is the case
and the ground tensor glueball is $g_T(2050)$, the mass of scalar
glueball should be $1.73\times \frac{2.05}{2.40}Gev=1.48Gev$,
which looks very like $f_0(1500)$. If the ground scalar glueball
is $f_0(1500)$, the dominant component should be $s\bar{s}$ in
$f_0(1710)$. This interpretation is favored in many phenomenon
analysis. For instance, we obtain a more natural conclusion that
the $s\bar{s}-n\bar{n}$ mass difference is  about $300MeV$\ct{q7}.
In fact, a more recent work\ct{0510066} reveals that one can not
distinguish the scalar glueball interpretation of $f_0(1500)$ and
$f_0(1710)$ nowadays in unquenched lattice QCD. From the above
argument we also obtain the mass of $1^{--}$ glueball should be
$3.85\times \frac{2.05}{2.40}Gev=3.29Gev$. This means that
$1^{--}$ glueball mass and $J/\psi$ mass are very close.
Therefore, taking account into the lattice computing error, to
solve the "$\rho-\pi$ puzzle" of $J/\psi$ and $\psi^\prime$
decays, the suggestion\ct{puzzle,puzzle1} that $J/\psi\rightarrow
\rho\pi$ is enhanced by a mixing of the $J/\psi$ with an $1^{--}$
glueball that decays to $\rho\pi$ is also suitable.

Unfortunately, in simulations we found that the mass of first
excited tensor glueball and the mass of second excited tensor
tensor glueball are very close. Such unexpected occasion makes the
traditional approach of variation principle, illustrated by C.
Morningstar\ct{prd97}, has disadvantages in the calculation. We
will show an alternative method in the future.

In summary, we have finished the first two steps in the
calculation of glueball spectrum. The leavings will be shown in
the future. Although we do not make any lattice simulation here,
we obtain an interesting conclusion, that is, $f_J(2220)$ and
$g_T$ states can not be tensor glueball simultaneously.

\end{document}